\documentclass[reprint, english,aps,prl,bibnotes]{revtex4-1}
\usepackage{graphicx}
\usepackage[T1]{fontenc}
\usepackage[latin9]{inputenc}
\usepackage{amstext}
\usepackage{amssymb}
\usepackage{multirow}
\usepackage{dcolumn}
\usepackage{color}
\usepackage{amsmath}
\usepackage{babel}
\makeatletter

\begin{document}

\title{Preserving entanglement during weak measurement \\ demonstrated with a violation of the Bell-Leggett-Garg inequality}

\author{T.C. White$^1$}
\thanks{These authors contributed equally to this work}
\author{J.Y. Mutus$^{1,4}$}
\thanks{These authors contributed equally to this work}
\author{J. Dressel$^{3}$}
\author{J. Kelly$^1$}
\author{R. Barends$^{1,4}$}
\author{E. Jeffrey$^{1,4}$}
\author{D. Sank$^{1,4}$}
\author{A. Megrant$^{1,2}$}
\author{B. Campbell$^1$}
\author{Yu Chen$^{1,4}$}
\author{Z. Chen$^1$}
\author{B. Chiaro$^1$}
\author{A. Dunsworth$^1$}
\author{I.-C. Hoi$^1$}
\author{C. Neill$^1$}
\author{P.J.J. O'Malley$^1$}
\author{P. Roushan$^{1,4}$}
\author{A. Vainsencher$^1$}
\author{J. Wenner$^1$}
\author{A. N. Korotkov$^{3}$}
\author{John M. Martinis$^{1,4}$}
\email{martinis@physics.ucsb.edu}
\affiliation{$^1$Department of Physics, University of California, Santa Barbara, California 93106-9530, USA}
\affiliation{$^2$Department of Materials, University of California, Santa Barbara, California 93106, USA}
\affiliation{$^3$Department of Electrical and Computer Engineering, University of California, Riverside, California, USA}
\affiliation{$^4$Present address: Google Inc., Santa Barbara, CA 93117, USA}

\date{\today}
\begin{abstract}

Weak measurement has  provided new insight into the nature of quantum measurement, by demonstrating the ability to extract average state information without fully projecting the system.  For single qubit measurements, this partial projection has been demonstrated with violations of the Leggett-Garg inequality. Here we investigate the effects of weak measurement on a maximally entangled Bell state through application of the Hybrid Bell-Leggett-Garg inequality (BLGI)  on a linear chain of four transmon qubits.  By correlating the results of weak ancilla measurements with subsequent projective readout, we achieve a violation of the BLGI with 27 standard deviations of certainty.
  
\end{abstract}
\maketitle

\begin{figure}
  
  \centering
    \includegraphics[width=0.45\textwidth]{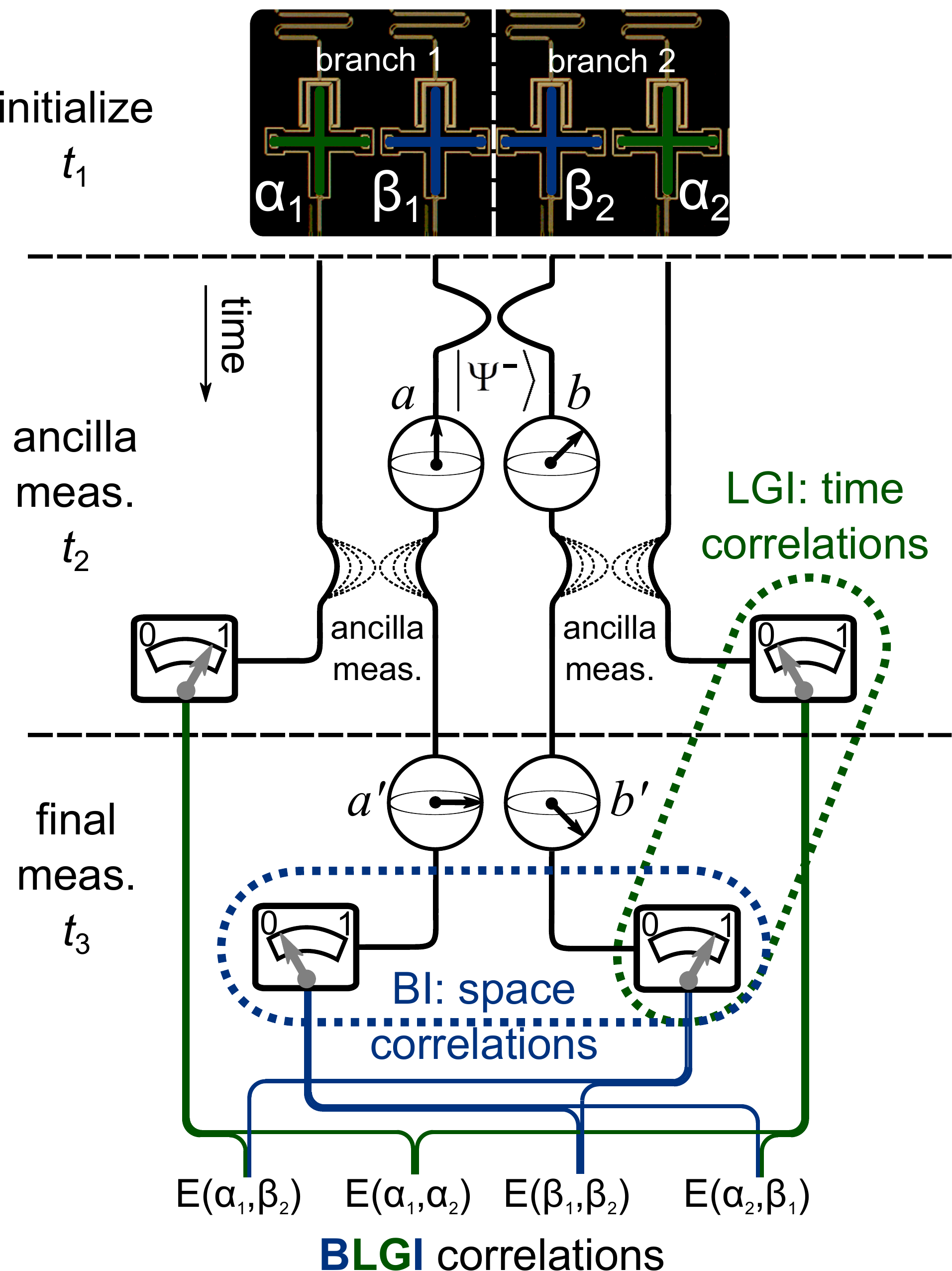}
  
\caption{Schematic of the hybrid Bell-Leggett-Garg inequality and optical micrograph of the superconducting quantum device.  The algorithm consists of two LGI weak measurement branches, bridged by the entanglement of the central Bell qubits.  The Bell pair ($\beta_{1,2}$) is initially prepared in the anti-symmetric singlet Bell state $|\Psi^- \rangle$.  Next, each Bell qubit is rotated to its first measurement basis ($a$ or $b$) and entangled with its ancilla qubit ($\alpha_{1,2}$). Finally, the Bell qubits are rotated and projectively read out in bases corresponding to angles $a'$ and $b'$.  By correlating the final projective read out and the weak ancilla measurements we calculate all four terms of a CHSH correlator simultaneously.}
\label{fig:inequalities}
\end{figure}

Quantum computing promises greater processing power through the clever application of superposition and entanglement.  Despite the importance of this uniquely quantum behavior, it occurs elusively behind the non-unitary effects of measurement collapse.  Weak measurements\cite{Kraus:weakMeas, aharonov:weakMeas, hatridge:weakMeas} have provided new insight into this process by demonstrating the ability to extract average state information without fully collapsing the system.  These gentler measurements have allowed single-configuration violations of the Leggett-Garg inequality \cite{ruskov:weakMeas,jordan:weakLGI,williams:weakLGI,palacios:weakmeasLG,goggin:opticalLG1,dressel:LGI,groen:ancillaMeas,emary:LGIreview} and, more recently, the detailed tracking of single qubit trajectories \cite{murch:trajectories,weber:mapping}. It is an outstanding challenge however, to achieve the same level of measurement control with an entangled state.  Here we demonstrate a continuous and controlled exchange between extracted single qubit state information and two-qubit entanglement collapse, through the unique framework of the Bell-Leggett-Garg inequality (BLGI).  We quantify this effect by correlating variable strength ancilla qubit measurements with subsequent projective readout to collect all the statistics of a Bell inequality experiment \cite{origBell, origCHSH,aspect:bell,bellReview} in a single quantum circuit.  Additionally, we demonstrate the ability to measure the Bell state with minimal entanglement collapse, by violating this hybrid BLGI\cite{dressel:loopholes} at the weakest measurement strengths.  This experiment indicates that it is possible carry out high fidelity ancilla measurement in large entangled systems. Additionally, combining this experiment with remote entanglement methods\cite{roch:remoteEntangle} may eventually lead to a loophole-free violation of classical hidden variable theories.

The challenge of successfully implementing weak measurements is twofold: the first is to evaluate the amount of information extracted on average by the measurement; the second is to evaluate the measurement back-action on the system.  For a single qubit state, the Leggett-Garg inequality \cite{origLG} (LGI)  provides an elegant way to do both with a single experiment.  The LGI was originally designed to verify the ``quantumness'' of macroscopic objects through the effects of projective measurement, which allows larger correlations between successive measurements (e.g., at times $t_1 < t_2 < t_3$) than are possible classically. More recent generalizations of the LGI prepare a known state at time $t_1$ and replace the intermediate measurement at time $t_2$ with a weak measurement \cite{ruskov:weakMeas,jordan:weakLGI,williams:weakLGI,palacios:weakmeasLG,goggin:opticalLG1,dressel:LGI,groen:ancillaMeas,emary:LGIreview}. This minimizes the quantum state disturbance while still extracting sufficient information on average.  Ideally this allows all the statistics necessary for a violation of the inequality to be measured with a \emph{single} experimental configuration.  Violating the inequality in this way guarantees that the state information has been extracted without significant back-action on the system \cite{palacios:weakmeasLG}.

Evaluating the effect of weak measurements on an entangled state is more difficult because the degree of entanglement is in generally challenging to quantify. The most robust method for quantifying entanglement remains a Bell test \cite{origBell}, which was first proposed by Bell and later refined by Clauser, Horne, Shimony, and Holt into an inequality (CHSH). The CHSH term sums the correlation measurements of two spatially separated qubits in four different measurement bases and bounds the maximum total value of classical correlations to be $|\text{CHSH}|_{\text{class}} \leq 2$.

In superconducting qubits, we use qubit state rotations to map the desired measurement basis onto the ground ($|0\rangle$) and excited ($|1\rangle$) states of the system.  For measurement rotations $a$ (qubit 1) and $b$ (qubit 2), the correlation amplitude between two measurements is given by
\begin{equation}
E(a,b) = P(00)-P(10)-P(01)+P(11),
\label{eq:E}
\end{equation}
where the term $P(00)$ is the probability both qubits are in the ground state. A traditional CHSH experiment combines the four correlator terms
\begin{equation}
\text{CHSH} = E(a,b) + E(a',b) + E(a,b') - E(a',b').
\label{eq:chsh}
\end{equation}
Entangled quantum states can violate the classical bound, with a fully entangled Bell
state ideally saturating the quantum upper bound of $|\text{CHSH}|_{\text{quant}} \leq 2\sqrt{2}$ at the specific rotation angles $a = 0$, $b = \pi/4$, $a' = \pi/2$, and $b' = 3\pi/4$.

To understand the effect of weak measurement on an entangled state we combine the spatial correlations of a Bell inequality with the temporal correlations of an LGI to construct a Bell-Leggett-Garg inequality (BLGI).  The algorithm, as described by Dressel and Korotkov \cite{dressel:loopholes} and shown in Fig.\,\ref{fig:inequalities}, consists of a CHSH-style experiment in which each Bell qubit is measured twice in succession as for a simultaneous LGI \cite{marcovitch:BLGI, palacios:weakmeasLG, groen:ancillaMeas}.  The initial measurements are carried out by ancilla qubits which act as probes of the entangled system before being projectively read out.  By varying the degree of entanglement between each ancilla and the Bell qubit it probes, we can vary the strength of the measurement. After preparing the Bell qubits in the anti-symmetric singlet state
\begin{equation}
|\Psi^- \rangle = \frac{1}{\sqrt{2}}(|01\rangle - |10\rangle),
\end{equation} 
each Bell qubit ($\beta_{1,2}$) is rotated to its first measurement angle ($a=0$, $b=\pi/4$) and then entangled with its ancilla qubit ($\alpha_{1,2}$) to implement the tunable-strength measurement. Next, each Bell qubit is rotated to its final measurement angle ($a'=\pi/2$, $b' = 3\pi/4$), and all four qubits are read out.  With this procedure the data for each measurement angle is encoded on a distinct qubit ($a \rightarrow \alpha_1$, $b \rightarrow \alpha_2$, $a' \rightarrow \beta_1$, and $b' \rightarrow \beta_2$). The BLGI correlator then takes the form similar to Eq.~\eqref{eq:chsh}
\begin{equation}
\langle C \rangle = -E(\alpha_1,\alpha_2) - E(\alpha_1,\beta_2) + E(\beta_1,\alpha_2) - E(\beta_1,\beta_2).
\end{equation}
where each term is calculated as in Eq.~\eqref{eq:E}.

The BLGI bounds correlations between these four distinct measurements based on the classical assumptions of local-macro-realism \cite{dressel:loopholes,supp}.  Classically the measurement of one qubit in the Bell pair should have no affect on the other, and the strength of the ancilla measurement should have no effect on the result of the following projective readout.  By encoding the measurement result for each rotation angle on an independent qubit, we can test both of these assumptions at the same time.  Quantum mechanically if the ancilla measurement is fully projective then we should measure the expected correlation amplitude of $|E(\alpha_1,\alpha_2)| = -1/\sqrt{2}$ only in the initial measurement basis, as the Bell qubits are no long entangled after that measurement.  As the ancilla measurement strength is decreased, we should extract the same qubit information on average while only partially collapsing the Bell state.  For sufficiently weak measurements, the magnitudes of all four correlators should approach the unperturbed Bell state values of $1/\sqrt{2}$ while $\langle C \rangle$ approaches $2\sqrt{2}$.  Thus a violation implies that our system demonstrates non-classical correlations through the entanglement of the Bell pair, while also demonstrating the ability to extract average state information without significant back-action on the entanglement of the system.

\begin{figure}

  \centering
    \includegraphics[width=0.45\textwidth]{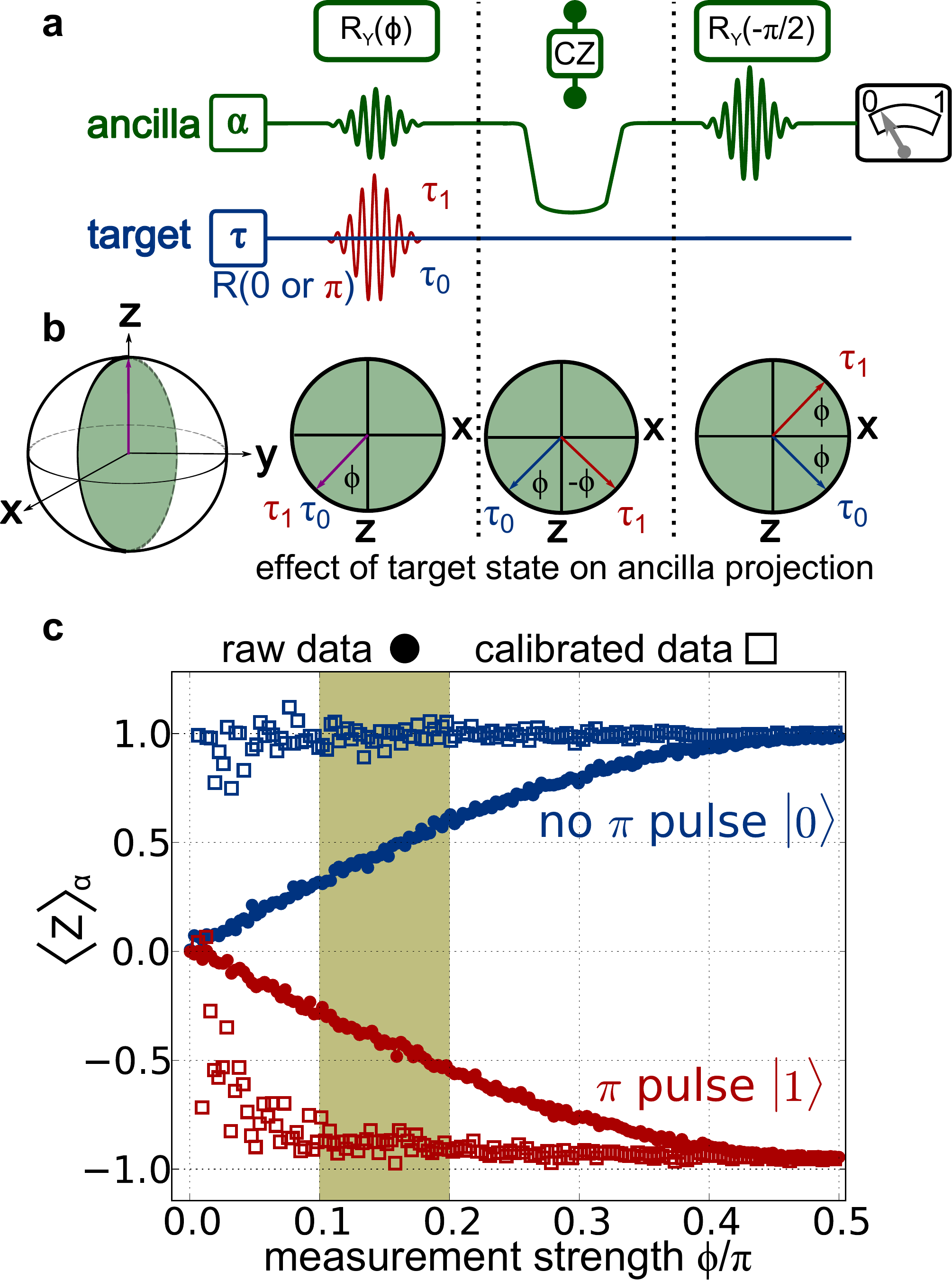}
 
  \caption{Weak measurement protocol. \textbf{a}, Full pulse sequence of the ancilla measurement algorithm used in the BLGI experiment. The measurement consists of a variable amplitude $Y$ rotation by an angle $\phi$ which controls the strength of the measurement.  This is followed by a CZ gate that entangles the ancilla qubit with the target qubit.  Finally the ancilla is rotated by an angle $-\pi/2$, bringing it into the desired measurement basis.  Two cases are compared, that of the target qubit in the ground (blue) or excited (red, $\pi$ rotation) state.  \textbf{b}, Bloch sphere representation of the ancilla qubit during the weak measurement protocol when the target qubit is in either the ground (blue) or excited (red) state.  The $Z$ averages of the ancilla and target qubit are correlated such that $\langle Z \rangle_\alpha = \sin(\phi)\langle Z \rangle_\tau$, where a full projective measurement corresponds to $\phi = \pi/2$ and no measurement corresponds to $\phi=0$. \textbf{c}, Ancilla measurement of prepared target state before and after calibrating for measurement strength.  We calibrate both curves by the scaling factor required to normalize the average 0 state curve.  This is almost equivalent to dividing by $\sin(\phi)$ but bounds the calibrated mean by $\pm1$.  In the calibrated case, the measured mean remains unchanged while the measured variance increases as $\phi$ decreases.  The gold shaded region denotes angles at which weak measurement data can violate the BLGI while still being reliably calibrated}
  
 \label{fig:weakMeas}
\end{figure}

We performed this experiment on a linear chain of Xmon transmon qubits, shown at the top of Fig.\,\ref{fig:inequalities}, with ground to excited state transition frequencies in the 4-6\,GHz range \cite{9xMon}. Each qubit is individually addressed with a microwave control line which can be used for single qubit $X$ or $Y$ gates as well as a DC line for implementing $Z$-gates and frequency control.  These control lines are used in conjunction to execute high fidelity two-qubit gates \cite{barends:gates} for entanglement and ancilla measurement.  The state of each qubit is measured independently using the dispersive shift of a dedicated readout resonator.  Resonators are frequency multiplexed \cite{chen:multiplexed} and readout with a broadband parametric amplifier \cite{IMPA}, which allows for fast high fidelity measurement.  Further details of this device can be found in Refs.\,\onlinecite{9xMon, supp}.

The ancilla measurement protocol used in this experiment and shown in Fig.~\ref{fig:weakMeas}, is a modified version of the protocol demonstrated in an LGI violation from Groen \textit{et al.} \cite{groen:ancillaMeas}.  Initially, an ancilla qubit is $Y$-rotated by an angle $0 \leq \phi \leq \pi/2$ from its ground state to set the measurement strength.  A control phase gate is then performed, causing a $Z$ rotation of $\pi/2$ in the ancilla qubit depending on the target qubit's state.  Finally a $-Y$ rotation of $\pi/2$ is performed on the ancilla qubit to rotate into the correct measurement basis.  The visibility of this measurement is then proportional to the distance of the ancilla state vector from the equator of the Bloch sphere, as shown in Fig.\,\ref{fig:weakMeas} (b).  When $\phi = \pi/2$ this operation becomes a control-NOT gate and implements a projective measurement. As $\phi \to 0$ the ancilla states become degenerate and no information is extracted.

As the final position of the ancilla state is dependent on the measurement strength, the ancilla readout is imperfectly correlated with the target qubit. That is, the visibility of an ancilla $Z$ average, $\langle Z \rangle_\alpha \simeq \sin(\phi) \langle Z \rangle_\tau$, is compressed from the target $Z$ average by a factor of approximately $\sin(\phi)$. To reconstruct the target $Z$ average from the ancilla $Z$ average, we should thus rescale the signal by $1/\sin(\phi)$.  Initially, this was done by a linear fit of the rotation angle $\phi$ to the qubit drive amplitude.  Unfortunately this linear fit was too rough to properly calibrate the smallest drive amplitudes and led to a systemic offset in the calibration of $\langle Z \rangle$. To keep the model as simple as possible, we kept the linear fit but used a data-based rescaling to set the measured ground state ($|0\rangle$) average to $1$, which ensures that the calibrated ancilla average is properly bounded by $\pm1$, as shown in Fig.\,\ref{fig:weakMeas}(c).  Further details of this calibration can be found in the supplemental information \cite{supp}.

\begin{figure}
  
  \centering
    \includegraphics[width=0.45\textwidth]{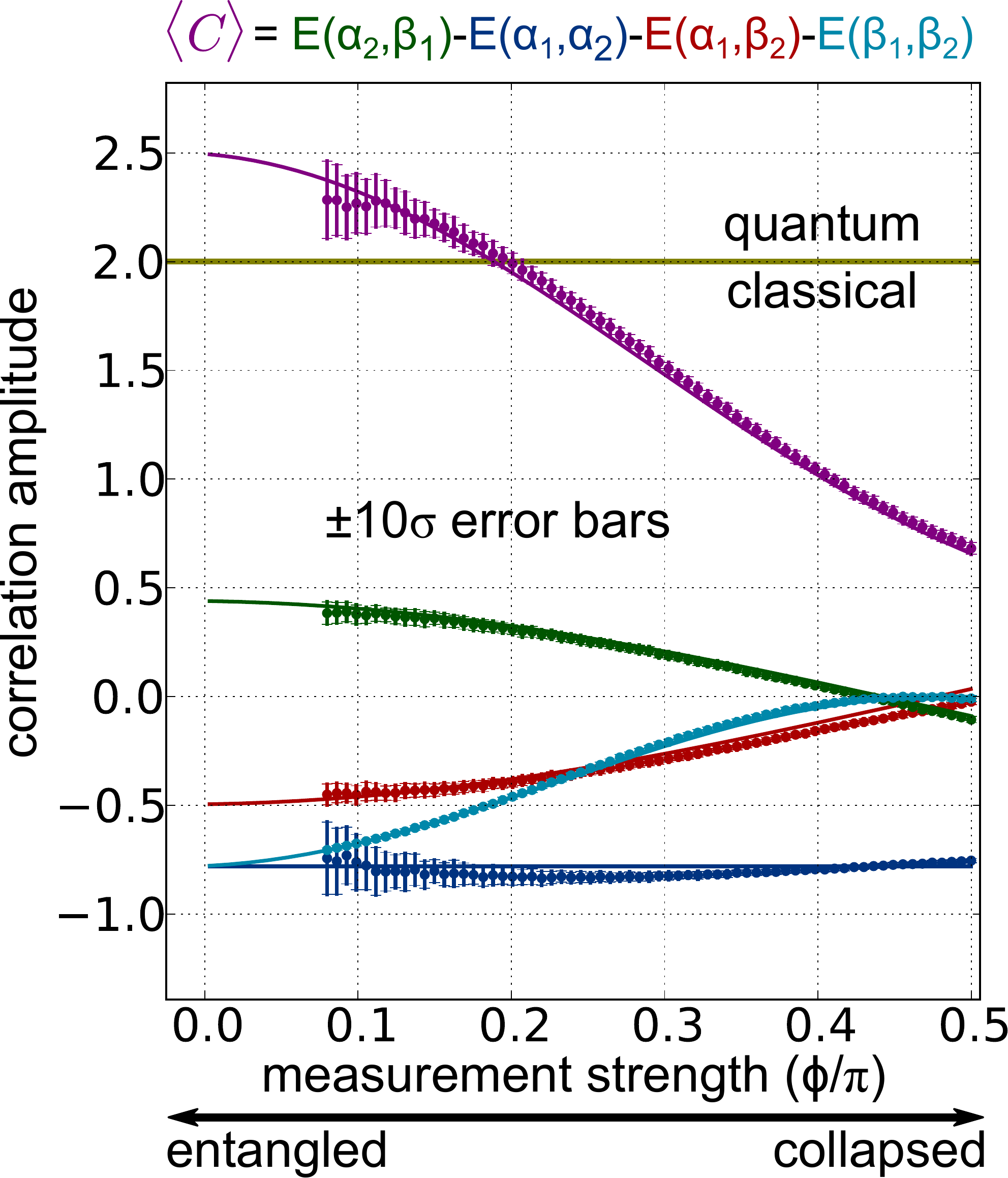}
\caption{Graph showing both experimental data (points) and theoretical predictions (lines) for the correlator $\langle C \rangle$ and its four terms vs. measurement strength $\phi$.  The horizontal gold line denotes the classical bound on $\langle C \rangle$.  The data set was taken by averaging together 200 traces in which each point was measured 3000 times for a total of 600,000 iterations per point.  The error bars represent 10 standard deviations of the mean to demonstrate the scaling the ancilla measurement noise vs. $\phi$.  The magnitude of the correlations between each pair of qubits reveals the extent to which entanglement has been broken for each measurement strength.}
  \label{fig:violation}
\end{figure}

After this and other extensive system calibrations, we can begin our investigation in earnest.  To quantify the entanglement collapse, we measure each two-qubit correlator in $\langle C \rangle$ vs. ancilla measurement strength $\phi$.  The data are plotted in Fig.\,\ref{fig:violation} alongside theory curves generated by a quantum model that includes realistic environmental dephasing and readout fidelity \cite{dressel:loopholes}.  Error bars for the data represent $\pm10$ standard deviations of the mean to demonstrate the increase in noise with decreasing measurement strength.  For projective angles $\phi\approx\pi/2$, the ancilla measurement results ($E(\alpha_1,\alpha_2)$) reflect the correlation expected from a fully collapsed Bell pair.  As measurement strength is decreased, this ancilla correlation remains nearly constant while additional inter-qubit correlations ($E(\alpha_1,\beta_2),$ $E(\beta_1,\alpha_2),$ $E(\beta_1,\beta_2)$) emerge.  For sufficiently weak measurements, $\langle C \rangle$ exceeds the classical bound of 2 and saturates towards the CHSH value of 2.5, which is expected from simulations for a fully entangled Bell state in realistic experimental conditions \cite{supp}.

The measured BLGI correlations follow the theoretical model \cite{supp} very closely for all measurement strengths.  This behavior reveals the continuous and controlled exchange between the collapse of an entangled Bell state and the single qubit state information gained from tunable-strength measurements.  Each ancilla qubit, when calibrated, retains the same correlations for all measurement strengths, whereas each Bell qubit has its correlations damped through partial projection by its ancilla qubit \cite{dressel:loopholes}.  The effect of partial projection can be seen in the difference in functional behavior between the Bell-ancilla ($E(\alpha_1,\beta_2),$ $E(\beta_1,\alpha_2)$) and the Bell-Bell ($E(\beta_1,\beta_2)$) correlator terms.  In the Bell-ancilla terms, the correlations are suppressed solely due to the randomization of the Bell qubit, but return as soon as measurement strength is  decreased.  In the Bell-Bell term, this effect is compounded as both qubits are being damped by partial ancilla projection, so the correlations return more slowly.  This gives $E(\beta_1,\beta_2)$ its distinct shape compared to the other correlators.  The BLGI terms can be seen in greater detail in the supplement \cite{supp}.

While other methods exist to characterize entanglement, the correlation measurements of a CHSH experiment remain one of the more robust tests of quantum behavior due to the considerations given to experimental loopholes.  Fortunately, the unique construction of the BLGI allows us to avoid some of the more pervasive loopholes appearing in traditional Bell or LG inequalities. The simultaneous measurement of all four CHSH terms in a single circuit allows us to avoid any configuration-dependent bias, such as the disjoint sampling loophole \cite{larsson:disjointLH}. The near unit detection efficiency in superconducting systems \cite{ansmann:CHSH,supp} similarly bypasses the fair sampling loophole \cite{pearle:fairsampLH}, which has hindered the investigation of related hybrid inequalities in optical systems \cite{dressel:LGI, higgins:blgi}.  Additionally, since the data from each ancilla qubit is only correlated with the data from the Bell and ancilla qubits on the remote LGI branch, we can substantially relax the usual LGI noninvasive measurement assumption to the standard locality assumption needed for a Bell inequality instead.

This locality assumption, fundamental to any Bell inequality, presumes no classical interactions between remote qubits occur during the correlation measurements.  The close proximity of adjacent superconducting qubits on a chip implies that such an interaction cannot be ruled out here.  Thus, behavior that appears to be quantum could, at least in principle, be the result of a fast classical interaction between hidden variables in the system.  While we cannot yet completely rule out these local interactions, there are several promising approaches to closing this loophole as well.  The assumptions of the BLGI requires only spatial separation of the central Bell qubits.  The speed and fidelity of operations in a superconducting qubit system makes modest spatial separation sufficient, and we can sacrifice some Bell state preparation fidelity to achieve it.  Techniques such as remote entanglement through measurement \cite{roch:remoteEntangle}, may soon provide the spatial separation necessary to conduct a loophole-free BLGI experiment.

Despite the few remaining loopholes, the excellent agreement between data and theoretical predictions in this experiment allows us to draw certain likely conclusions about the application of ancilla measurement in superconducting circuit systems.  The functional dependence of the BLGI correlator on measurement strength implies that the back-action is dominated by the projectiveness of the measurement.  This makes it unlikely there is some poorly understood classical error mechanism which would make it difficult to implement error correction schemes such as the surface code \cite{fowler:surface}, which rely on sequential high fidelity ancilla measurements to detect errors \cite{9xMon}.  The violation of the BLGI at the weakest measurement strength  also implies that it is possible to extract average state information without significant back-action. Normally the usefulness of this type of measurement is limited by the large number of statistics required to average the noisy detector output and the coherence time of the qubits limiting the total number of measurements.  It should be possible however, to integrate weak ancilla measurements into a large surface code cell, which would correct for most errors while also allowing for a larger ensemble to be collected from the weak measurement.  Thus weak ancilla measurement could become an important tool in understanding the dynamics of large quantum systems.

In the course of this work we have demonstrated the continuous and controlled collapse of an entangled state based on the strength of tunable ancilla measurements.  This behavior was quantified using the simultaneous correlation measurements which make up the Bell-Leggett-Garg inequality.  The violation of this inequality at the weakest measurement strengths demonstrates the viability of using weak ancilla measurements to conduct many sequential measurements of entangled states.  This provides a window into the evolution of entangled states, a critical component in scaling to larger quantum systems.  With the inclusion of new remote entanglement algorithms the BLGI may also lead to a loophole-free violations of classical hidden variable theories. Lastly this demonstrates that as we scale to larger multi-qubit systems, with the fidelity and control achieved here, we gain greater access to the rich physics at the heart of quantum mechanics.

\textbf{Acknowledgments} This work was supported by the Office of the Director of National Intelligence (ODNI), Intelligence Advanced Research Projects Activity (IARPA), through the Army Research Office grant W911NF-10-1-0334. All statements of fact, opinion or conclusions contained herein are those of the authors and should not be con-strued as representing the official views or policies of IARPA, the ODNI, or the U.S. Government.  Devices were made at the UC Santa Barbara Nanofabrication Facility, a part of the NSF-funded National Nanotechnology Infrastructure Network, and at the NanoStructures Cleanroom Facility. 

\textbf{Author contributions} T.C.W. and J.Y.M. performed the experiment and analyzed the data. J.D. designed the experiment together with T.C.W., and J.Y.M.  J.K., A.E.M.,
and R.B. fabricated the sample.  T.C.W., J.Y.M., and J.M.M co wrote the manuscript.  All authors contributed to the fabrication process, experimental setup and manuscript preparation.

\textbf{Additional information} The authors declare no competing financial interests. Supplementary information accompanies this paper on [weblink to be inserted by editor]. Reprints and permissions information is available at [weblink to be inserted by editor]. Correspondence and requests for materials should be addressed to T.C.W., J.Y.M. or J.M.M


\begin{thebibliography}{10}
\bibitem{swapnote}
\bibinfo{note}{In refs. \cite{barends:gates} and \cite{kelly:orbit}, the
  convention was used where g/2$\pi$ is the swap rate. Here we use 2g/2$\pi$ as
  the swap rate, in accordance with others in the field.}
\end{thebibliography}

\end{document}


\title{Supplementary Information}

\date{\today}

\section{Supplemental information}

\subsection{Bell and Leggett-Garg Inequalities}

\begin{figure}[]
    \centering
    \includegraphics[width=0.95\textwidth]{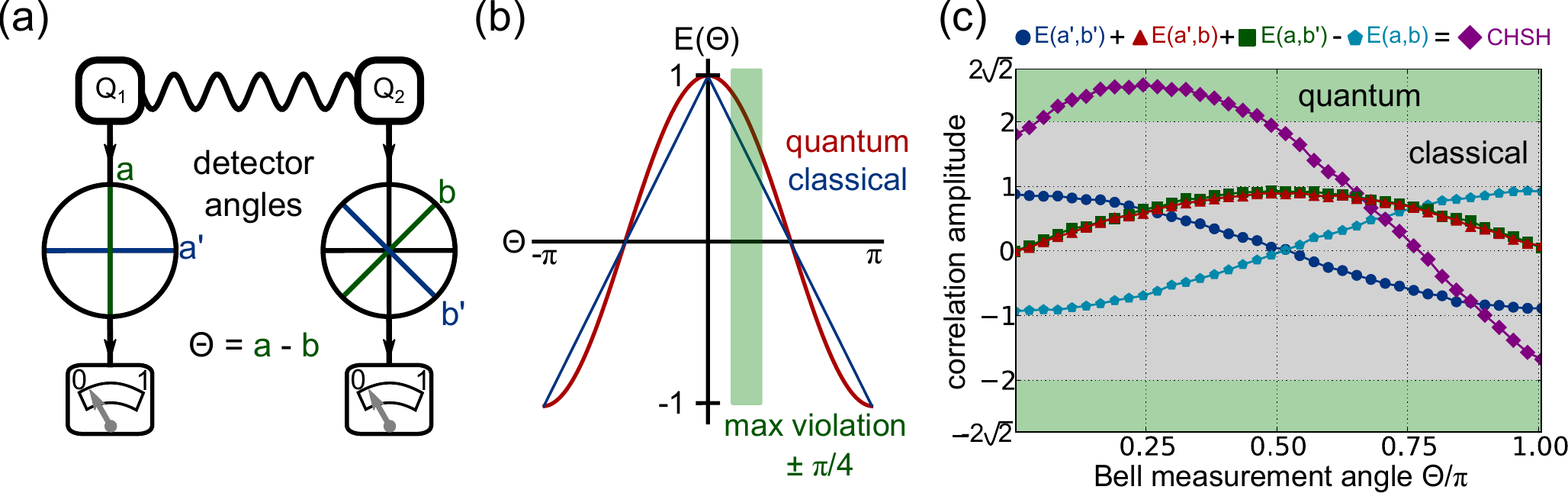}
    \caption{Entanglement and the CHSH corralator. (a) Schematic of a typical CHSH experiment.  Two qubits are prepared in one of the four possible Bell states and measurements are conducted at various measurement angles $a$, $b$, $a' = a+\pi/2$, and $b' = b+\pi/2$.  (b) Dependence of correlation amplitude (Eq.\,\ref{eq:E}) on detector angle difference $\theta = a-b$.  The difference between quantum and classical correlation is maximized at $\theta = \pi/4$.  (c) Measurement of CHSH correlator using the central Bell qubits prepared in a $|\phi^+ \rangle$ Bell state, displaying the sinusoidal dependence on the relative measurement angle $\theta$.  This experiment avoids common sampling loopholes with high detector efficiency, but is subject to the locality loophole due to the small spatial separation of adjacent superconducting qubits on a chip.}
    \label{fig:CHSH}
\end{figure}

The CHSH correlator, designed by Clauser, Horne, Shimony, and Holt\cite{origCHSH} as a refinement of the Bell inequality\cite{origBell}, provides a quantitative bound on classical hidden variable theories using correlated measurements between two spatially separated qubits.  The correlator combines four different experimental configurations because it can be difficult to tell the difference between potentially classical (un-entangled) qubits and an entangled state in only one basis.  With superconducting qubits, the measurement basis for each qubit is set using qubit rotations to map the desired state onto the ground ($|0\rangle$) and excited ($|1\rangle$) states of the system. For measurement rotations $a$ (qubit 1) and $b$ (qubit 2), shown in Fig.\,\ref{fig:CHSH}(a), the correlation amplitude is given by
\begin{equation}
E(a,b) = P(00)-P(10)-P(01)+P(11),
\label{eq:E}
\end{equation}
where P(00) is the probability both qubits are in the ground state. Given this equation we can see that both the Bell state $|\Phi^+ \rangle = (|00\rangle + |11\rangle)/\sqrt{2}$ and the prepared state $|00\rangle$ will have a correlation amplitude of 1 if $a=b=0$.  The difference only becomes clear when the detector angle of one qubit is rotated relative to the other.  The behavior of $E(a,b)$ vs detector rotation, described here as $\theta = a-b$, is shown in Fig.\,\ref{fig:CHSH}(b) for both the classical and quantum case.  If the two objects can be described separately, then $E$ is only a linearly dependent on $\theta$.  If the two objects are entangled, then $E$ is has a sinusoidal dependence on $\theta$ with the maximum difference occurring at $\theta = \pi/4$.  

To initially characterize the system we conducted a traditional CHSH experiment using the central Bell qubits ($\beta_{1,2}$).  The relative measurement angles for each qubit were held fixed such that $a' = a + \pi/2 $ and $ b' = b+ \pi/2$.  We then varied $\theta = a-b$ from 0 to $\pi$, and measured each individual correlator as well as the sum given by
\begin{equation}
\text{CHSH} = E(a,b) + E(a',b) + E(a,b') - E(a',b').
\end{equation}
For any two classical states measured at these angles, we should see a linear dependence of $E(\theta)$ and a bound on the the CHSH correlator of $|CHSH|\leq2$.  Alternately, if the two qubits are in a maximally entangled Bell state, we should see sinusoidal behavior for $E(\theta)$ and a maximum CHSH value of $2\sqrt{2}$.  The data, shown in Fig.\,\ref{fig:CHSH}(c), display the expected sinusoidal dependence for each individual term, with a maximum CHSH amplitude near $\theta = \pi/4$.  While this data shows a robust violation of the classical bound, it fails to reach the theoretical maxium bound of $2\sqrt{2}$.  The maximum CHSH amplitude of $\sim2.5$ we see here is due to experimental imperfections which will be discussed later.  This CHSH experiment provides the framework  for the BLGI, as well as a benchmark for the maxium violation we should expect at the weakest measurement angles.

A complementary test of quantum mechanics is the LGI, which is similar to a Bell inequality but involves measurements separated in time rather than in space.  Classical theories of measurement assume that the system is always in a definite state, and that an ideal measurement will not change the state of the system.  In contrast, if one were to measure a quantum state in an orthogonal measurement basis, the act of measurement would project that object onto an eigenstate of the new basis.  To distinguish one kind of system from the other, measurements are conducted in different bases at different times.  For measurements conducted at times $t_1<t_2<t_3$, we can construct correlators analogous to Eq.\,\ref{eq:E} but for different measurements of the same qubit,
\begin{equation}
E(t_i,t_j) = P(00)-P(10)-P(01)+P(11).
\end{equation}
The inequality was originally composed of three distinct experiments.  In the first experiment, the system is measured projectively at time $t_1$, followed by a final projective measurement at time $t_3$. A second experiment is then carried out where an intermediate measurement in a different basis is conducted at time $t_2$ instead of time $t_3$.  The third experiment consists of only the measurements at times $t_2$ and $t_3$. The LGI is then given by
\begin{equation}
-3 \leq E(t_1,t_2) + E(t_2,t_3) - E(t_1,t_3)\leq 1
\end{equation}
where $E(t_1,t_3)$ is the experiment in which no measurement is performed at time $t_2$.  Further details for LGIs can be found in the review article by Emary \textit{et al.} \cite{emary:LGIreview}.

The weak measurement techniques discussed in the main text were created to avoid the possibility of a ``clumsy'' measurement loophole \cite{wilde:clumsyLH}. When sequential measurements are performed on the same system, it is impossible to ensure that LGI violations are not due to overly invasive measurements perturbing the system in an unknown way.  To minimize the effect of measurement, most LGIs replace the measurement at time $t_1$ with preparation of a known state, and the measurement at time $t_2$ with a null \cite{knee:nullMeas} or weak \cite{palacios:weakmeasLG,groen:ancillaMeas} measurement. These weak measurements \cite{palacios:weakmeasLG,groen:ancillaMeas} minimize back action on the system, while still extracting enough information to identify its state.  Using this technique, all the statistics of the LGI can be measured by conducting all three ``measurements'' in a single experimental configuration.  To construct the Bell-Leggett-Garg inequality we combined a traditional CHSH experiment with this weak measurement technique.  This allows us to measure all four terms of the CHSH correlator simultaneously in a single experiment.

\subsection{BLGI Algorithm Assumptions and Loopholes}

The fundamental assumptions of the hybrid Bell-Leggett-Garg inequality are those of \emph{local realism}, which are familiar from the Bell inequalities:
\begin{enumerate}[(i)]
  \item If an object has several distinguishable physical states $\lambda$, then at any given time it occupies only one of them.
  \item A measurement performed on one object of a spatially-separated pair cannot disturb the second object.
  \item Measured results are determined causally by prior events.
\end{enumerate}

Note that only assumption (ii) differs from the notion of \emph{macrorealism} used in Leggett-Garg inequalities: it is weakened here to permit \emph{local} invasiveness for sequential measurements in time made on the same object, while still forbidding spatially \emph{remote} measurements from influencing each other. Note that the assumed physical state $\lambda$ may be related to the quantum state, or may be a collection of more refined (but unspecified) hidden variables.

To these core assumptions we must append one more to permit noisy (i.e., realistic) detectors:
\begin{enumerate}[(iv)]
  \item Unbiased noisy detectors produce results that are correlated with the true object state $\lambda$ on average.
\end{enumerate}
This assumption can be understood as follows. The object state $\lambda$ ideally determines each measurable property $A(\lambda)$, but a physical detector (and environment) that interacts with the system will also have a distinct physical state $\xi$ that may fluctuate noisily between realizations (e.g., from the coupling procedure). In such a case the detector will report a correspondingly fluctuating signal $\alpha(\xi)$ according to some response probability $P_A(\xi | \lambda)$ for obtaining the detector state $\xi$ given each definite system state $\lambda$. For any sensible detector, these response probabilities will be fixed by the systematic and repeatable coupling procedure (such as our ancilla measurement circuit). To calibrate such a detector, we must then assume that averaging over many realizations of the detector noise will faithfully reflect information about each \emph{prepared} system state $\lambda$ (even if that state ultimately changes for subsequent measurements due to the coupling):
\begin{align}\label{eq:noiseaverage}
  \textstyle{\sum_\xi}\, \alpha(\xi)\, P_A(\xi | \lambda) = A(\lambda).
\end{align}
Importantly, this equality formally states only what is usually assumed for an unbiased laboratory detector: that one can recover a meaningful system value $A(\lambda)$ by averaging away any detector noise.

Now consider the Bell-Leggett-Garg correlation. A correlated pair of objects with the joint state $\lambda$ is sampled from an ensemble with the distribution $P(\lambda)$. (In our experiment, we prepare two qubits in a Bell state.) At a later time each object ($k=1,2$) is coupled to a detector (an ancilla qubit) that outputs a noisy signal $\alpha_k$ calibrated to measure the bounded property $A_k(\lambda)\in[-1,1]$ on average (the $Z$ operator for each Bell qubit).  The noisy signal $\alpha_k$ generally has an expanded range of values that can lie outside the range $[-1,1]$ (in our case $\alpha_k \approx \pm 1/\sin\phi$); however, for each $\lambda$ the realizations of the output signal will average to the correct bounded value by assumption (iv).  (We verify this assumption with the ancilla calibration measurements using definite preparations of $0$ or $1$ on the Bell qubits.) Finally, each object is measured with a second detector that outputs a signal $b_k$ for a similarly bounded property $B_k(\zeta)\in[-1,1]$ (we read out the qubits directly to obtain $b_k = \pm 1$).  From these four measured numbers, we then compute the CHSH-like correlator as a single number for each preparation 
\begin{align}\label{eq:corr}
  C = -\alpha_1 \alpha_2 - \alpha_1 b_2 + b_1 \alpha_2 - b_1 b_2.
\end{align}

The expanded ranges of the noisy signals $\alpha_k$ generally produce a similarly expanded range for the correlator $C$ for each preparation.  Nevertheless, averaging $C$ over many realizations of the detector noise $\xi_{A_k}$ and $\xi_{B_k}$ and system states $\lambda$ will produce
\begin{align}
  \mean{C} &= \textstyle{\sum_\lambda \sum_{\substack{\xi_{A_1},\xi_{B_1} \\ \xi_{A_2},\xi_{B_2}}}}\, C\, P(\xi_{A_1},\xi_{B_1} | \lambda)P(\xi_{A_2},\xi_{B_2}|\lambda)P(\lambda), \nonumber \\
  &= \textstyle{\sum_\lambda} \, \big[-A_1(\lambda) A_2(\lambda) - A_1(\lambda) B'_2(\lambda) + B'_1(\lambda) A_2(\lambda) - B'_1(\lambda) B'_2(\lambda)\big]\, P(\lambda),
\label{eq:corrav}
\end{align}
with $A_k(\lambda) = \sum_{\xi_{A_k},\xi_{B_k}} \alpha_k(\xi_{A_k})\, P(\xi_{A_k},\xi_{B_k}|\lambda)$ and $B'_k(\lambda) = \sum_{\xi_{A_k},\xi_{B_k}} b_k(\xi_{B_k})\, P(\xi_{A_k},\xi_{B_k}|\lambda)$, since postulate (ii) causes the joint distribution of the detector states to factor: $P(\xi_{A_1},\xi_{B_1},\xi_{A_2},\xi_{B_2}|\lambda) = P(\xi_{A_1},\xi_{B_1}|\lambda)P(\xi_{A_2},\xi_{B_2}|\lambda)$ in the same way as for a Bell inequality.  From the postulates (i), (iii), and (iv), the averages $A_k(\lambda)$ and $B'_k(\lambda)$ are then bounded to the range $[-1,1]$.  Therefore, for each $\lambda$ the sum of the bounded averages in Eq.~\eqref{eq:corrav} must itself be bounded by $[-2,2]$.  Averaging this bounded result over $P(\lambda)$ produces the expected BLGI
\begin{align}
  -2 \leq \mean{C} \leq 2.
 \label{eq:chsh}
\end{align}

Importantly, the joint probability $P(\xi_{A_k},\xi_{B_k}|\lambda) = P(\xi_{A_k}|\lambda)P(\xi_{B_k}|\lambda,\xi_{A_k})$ for each qubit $k$ admits the dependence of the $B'_k$ measurement on an \emph{invasive} $A_k$ measurement that can alter the physical state $\lambda$.  Despite any randomization of the results $b_k(\xi_{B_k})$ caused by such local invasiveness, however, the perturbed averages $B'_k(\lambda)$ must still lie in the range $[-1,1]$ since each $b_k = \pm 1$ by construction.  This allowance for \emph{locally} invasive measurements in the BLGI is what avoids the clumsiness loophole \cite{wilde:clumsyLH} of the usual LGI.  The fact that the entire correlator $C$ is computed for every realization in the same experimental configuration is what avoids any variant of the disjoint sampling loophole \cite{larsson:disjointLH} for the usual Bell and LGIs (such as from systemic bugs in the preparation software).

There are, however, two notable ways that our derivation of the BLGI in Eq.~\eqref{eq:chsh} could fail.  First, the assumptions (i--iii) of local realism could fail, as in a standard Bell inequality.  This is certainly possible in our case since the Bell qubits are neighbors on the same superconducting chip.  However, arranging for a locally realist model that accounts for the needed disturbance effects for the neighboring Bell qubits, the neighboring Bell-ancilla qubits, each remote pair of Bell-ancilla qubits, and the remote ancilla-ancilla qubits simultaneously is substantially more difficult (and therefore much less likely) than arranging for such disturbance in the usual Bell test on just two neighboring qubits. Moreover, our experiment verifies the detailed functional dependence of the quantum predictions as the weak measurement angle $\phi$ is varied, which further constrains any purported locally realist explanation. Thus our tested BLGI significantly tightens the locality loophole \cite{larsson2014loopholes} compared to the usual Bell test performed on the same chip.

Second, the noisy detector assumption (iv) could fail due to hidden preparation noise $\xi_P$ not included in the state $\lambda$ that systematically affects the detector output in both arms in a correlated way. In this case, the detector response would become noise-dependent $P_A(\xi|\lambda) \to P_A(\xi|\lambda,\xi_P)$ such that the calibration of Eq.~\eqref{eq:noiseaverage} will be satisfied only after additionally averaging over $\xi_P$.  Such correlated noise would prevent the detector distributions from factoring for each $\lambda$ in Eq.~\eqref{eq:corrav}, which formally spoils the inequality.  However, in our experiment such a systematic bias due to correlated noise has been extensively checked during the characterization of the chip and the measurement calibration by deliberately preparing a variety of uncorrelated distributions $P(\lambda)$ (i.e., different qubit states) and looking for spurious cross-correlations of the various qubit readout signals that would be expected in the presence of such hidden preparation noise. Hence, the failure of assumption (iv) additionally requires an unlikely preparation-conspiracy where every calibration check that has been done is somehow immune to the hidden detector-noise correlations.

\subsection{Weak Measurement Calibration}

\begin{figure*}[t]
    \centering
    \includegraphics[width=0.95\textwidth]{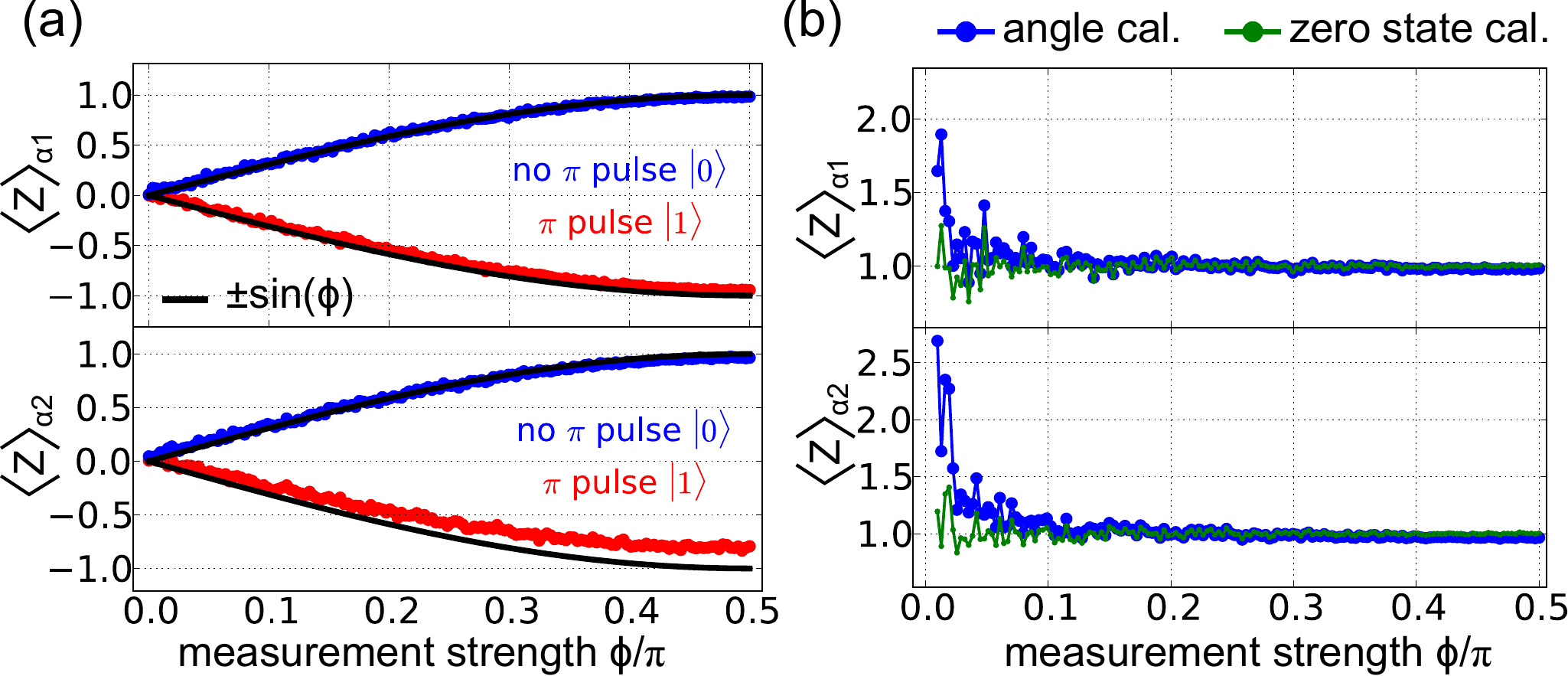}
    \caption{(a) Raw measurement of $\langle Z \rangle$ vs. measurement strength $\phi$ for both $\alpha_1$ and $\alpha_2$ along with an ideal curve of $\pm \sin(\phi)$.  Due to decoherence, measurement error, and possible calibration error the data does not perfectly follow the $\pm \sin(\phi)$ curves and converges to a value slightly below 0 at the weakest measurement strength. (b) $|0\rangle$ state calibrations done using both the $1/\sin(\phi)$ method (blue) and the point by point calibration (green).  In the $1/\sin(\phi)$ calibration the mean of the $|0\rangle$ state measurement drifts above 1 at the weakest measurement angles, while the mean follows 1 for the point by point case. }
    \label{fig:weakMeasCal}
\end{figure*}

As discussed in the main text the ancilla readout is imperfectly correlated to the Bell qubit's state.  When measuring in the $Z$ basis, $\langle Z \rangle_\alpha = \sin(\phi) \langle Z \rangle_\beta$, shown in Fig.\,\ref{fig:weakMeasCal} (a) along with the ideal curves $\pm \sin(\phi)$.  To calibrate this weak measurement we must first relate the measurement angle $\phi$ to microwave drive power, by fitting to a measurement of $|1\rangle$ state probability vs. $\pi$-pulse amplitude. The most straight forward calibration would then be to divide $\langle Z \rangle_\alpha$ by $\sin(\phi)$ shown in the blue curves in Fig.\,\ref{fig:weakMeasCal} (b), but this method causes drift in the mean at the smallest angles.  

This simple calibration fails because the raw data curves shown in Fig.\,\ref{fig:weakMeasCal} (a) converge to a value slightly below zero.  This means that for the weakest measurements, the simple angle calibration will under-correct a $|1\rangle$ state measurement and over-correct a $|0\rangle$ state measurement.  This over-correction of the $|0\rangle$ state is problematic, since calibrated values for $\langle Z \rangle_\alpha$  not bounded by $\pm 1$ will possibly violate the inequality incorrectly.  To prevent this, we instead use a data based calibration for each ancilla using the average of the $|0\rangle$ state measurement curve.  This has the advantage of bounding the mean of the calibrated result by $\pm 1$, at the expense of accentuating the drift in the mean of the $|1\rangle$ state measurement towards 0.  The results of this $|0\rangle$ state calibration are shown in the green curves in Fig.\,\ref{fig:weakMeasCal} (b).

To apply this calibration to the correlator terms, we must first express them in terms of the measurement operator $\langle Z \rangle$.  In a superconducting system the state rotations are used to map the desired measurement basis onto the ground ($|0\rangle$) and excited ($|1\rangle$).  For the ancilla measurement this is equivalent to mapping onto the $Z$ measurement axis.  Given probability P(1) of measuring the excited ($|1\rangle$) state, $\langle Z \rangle = 1-2P(1)$. After mapping state probabilities onto the $Z$ measurement axis we can express the correlator as $E(\alpha,\beta) = \langle Z \rangle_\alpha \langle Z \rangle_\beta$.  Expressed in this way we can see that for calibration factor $cal(\phi) \approx 1/\sin(\phi)$, $E_{cal}(\alpha,\beta) = E(\alpha,\beta)*cal(\phi)$.  Extending this to the BLGI we calibrate each term depending on the ancilla qubit being measured such that $E(\alpha_1,\beta_2)\rightarrow E(\alpha_1,\beta_2)*cal(\phi_1),$  $E(\beta_1,\alpha_2) \rightarrow E(\beta_1,\alpha_2)*cal(\phi_2),$ $E(\alpha_1,\alpha_2)\rightarrow E(\alpha_1,\alpha_2)*cal(\phi_1)*cal(\phi_2),$ and $E(\beta_1,\beta_2)$ remains unchanged.

\subsection{Error Analysis and Pulse Sequence Optimization}

\begin{figure}[b]
   \centering
    \includegraphics[width=0.95\textwidth]{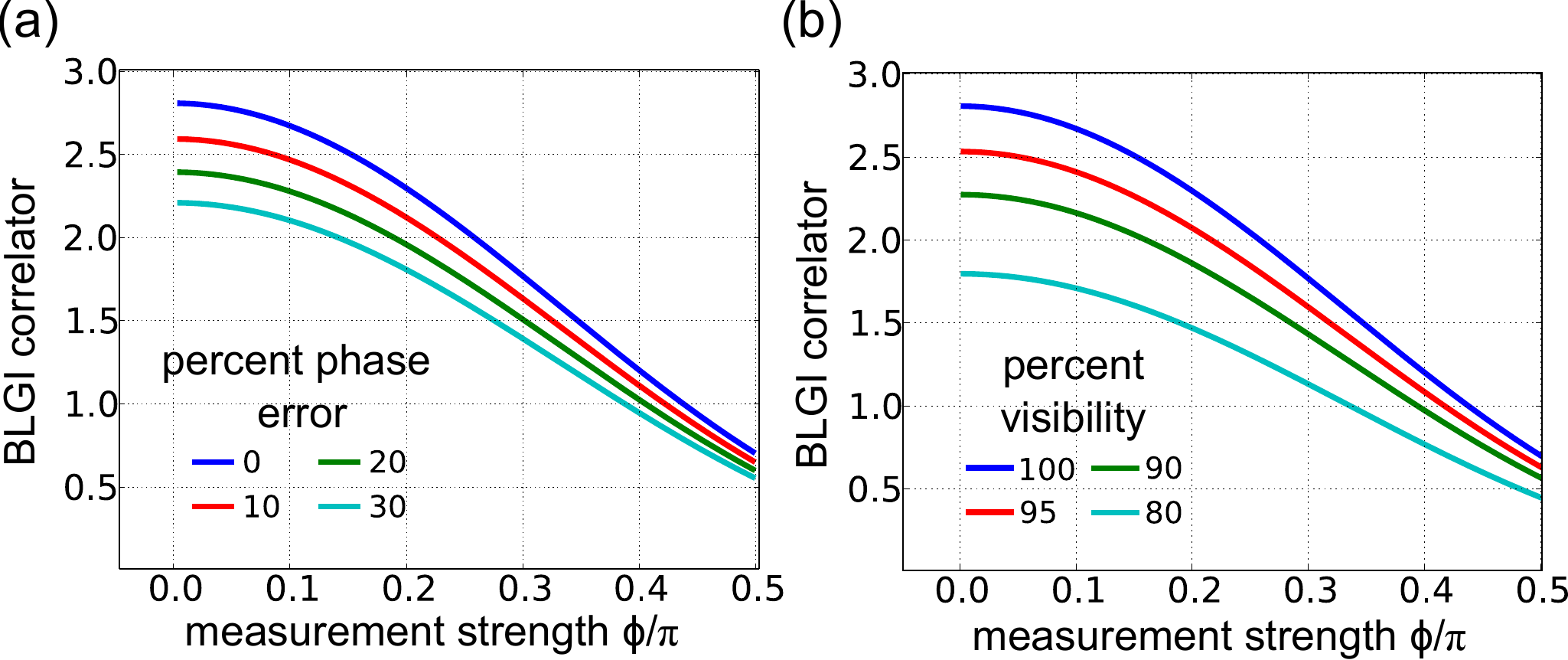}
    \caption{Effect of error mechanisms on BLGI correlations. (a) Amplitude of $\langle C\rangle$ vs. measurement strength for various single qubit dephasing error rates.  The single qubit dephasing error rate can be thought of as roughly $T_2$/(gate time) in the simplest case, but can be improved through the addition of echo pulses.  (b) Amplitude of $\langle C\rangle$ vs. measurement strength for various single qubit measurement visibility values.  The system shows a greater sensitivity to this error mechanism and cannot tolerate a visibility much below 90 percent.}
    \label{fig:errorEffects}
\end{figure}

While the algorithm and weak measurement scheme are simple in design, dependence on correlations between multiple qubits makes $\langle C \rangle$ sensitive to multiple error mechanisms.  As all four qubits were operated away from the flux insensitive point they were more susceptible to dephasing effects.  The Xmon qubits used in this experiment were extensively characterized in Ref. \cite{9xMon}, with a $T_2$ Ramsey decay of order $2-3\,\mu$s at the idle point.  When Characterizing the phase error per gate with randomized benchmarking, we find an error of roughly 0.25 percent per gate.  This corresponds to a $T_{2 \text{echo}}$ of over $10\,\mu$s.    The amplitude of $\langle C \rangle$ vs. dephasing error per qubit is shown in Fig.\,\ref{fig:errorEffects}(a).  The violation amplitude is relatively robust to this error, and can sustain error rates of up to 30 percent while still exhibiting non-classical correlations.  The second major error mechanism was reduced measurement visibility coming from $T_1$ energy decay or spurious $|1\rangle$ state population.  The effect on $\langle C \rangle$ vs. single qubit measurement visibility is shown in Fig\,\ref{fig:errorEffects}(b).  The correlation amplitude is more sensitive to this reduced measurement visibility and is significantly degraded at even 90 percent.  In both cases, the presence of errors not only lowers the maximum violation possible but the highest measurement strength at which a violation first occurs.  As weaker measurement angles require finer calibration and provide noisier data, it is preferable to achieve a violation at the largest measurement strength possible.

\begin{figure}[t]
    \centering
    \includegraphics[width=0.5\textwidth]{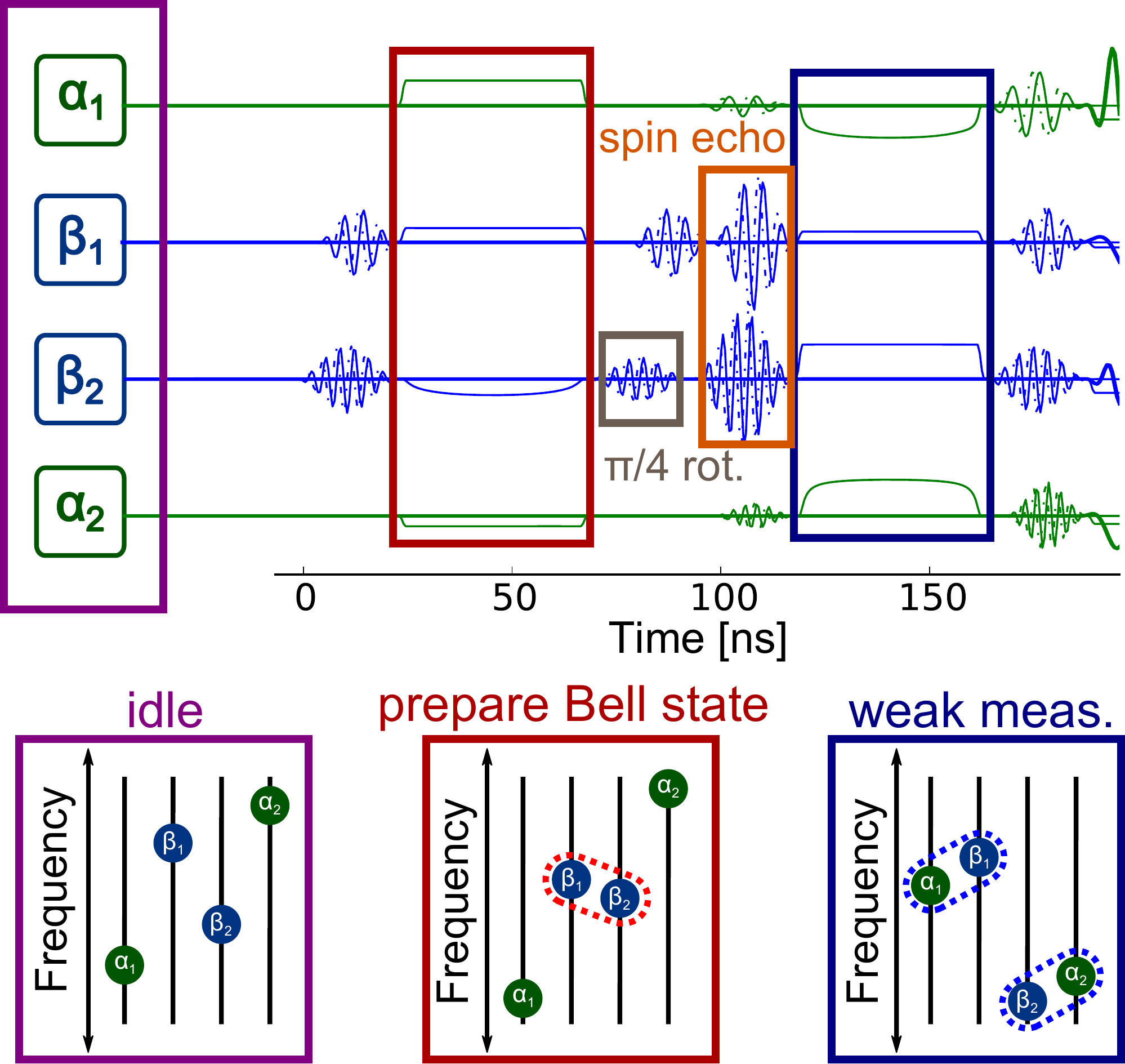}
    \caption{(above) Full pulse sequence of the BLGI algorithm, including spin echo pulses on the Bell qubits to reduce dephasing.  The weak measurements were carried out simultaneously to further streamline the algorithm.  (below) Frequency diagram of the 4 qubits showing their placement in frequency space the pulse sequence, in particular the various detunings during the adiabatic $CZ$ gates.}
    \label{fig:pulseSeq}
\end{figure}

Given the sensitivity of $\langle C\rangle$ to various decoherence mechanisms it was important to reduce the BLGI pulse sequence time as much as possible for higher coherence.  This is most notable during the weak measurement portion when we carry out simultaneous adiabatic CZ gates\cite{barends:gates} between both ancilla-Bell pairs.  Lastly, we introduced spin echo pulses in the middle of the algorithm which cancel out dephasing during the pulse sequence while simply transforming the original $|\Psi^{+}\rangle$ Bell state to a $|\Psi^{-}\rangle$.  To maximize measurement fidelity, we used a wide bandwidth parametric amplifier \cite{IMPA}, to ensure a high signal to noise ratio and shorter readout time. A separate measurement at the beginning of the pulse sequence was used to herald \cite{johnson:heralded} the qubits to the ground state, but this was a small ($\sim$6 percent) effect.  Lastly, we implemented numeric optimization of the adiabatic $CZ$ gates using the ORBIT protocol \cite{kelly:orbit} to fine tune parameters for the final data set.  The full Pulse sequence and frequency placement of the qubits during the algorithm is shown in Fig.\,\ref{fig:pulseSeq}.

\begin{figure}[t]
    \centering
    \includegraphics[width=0.5\textwidth]{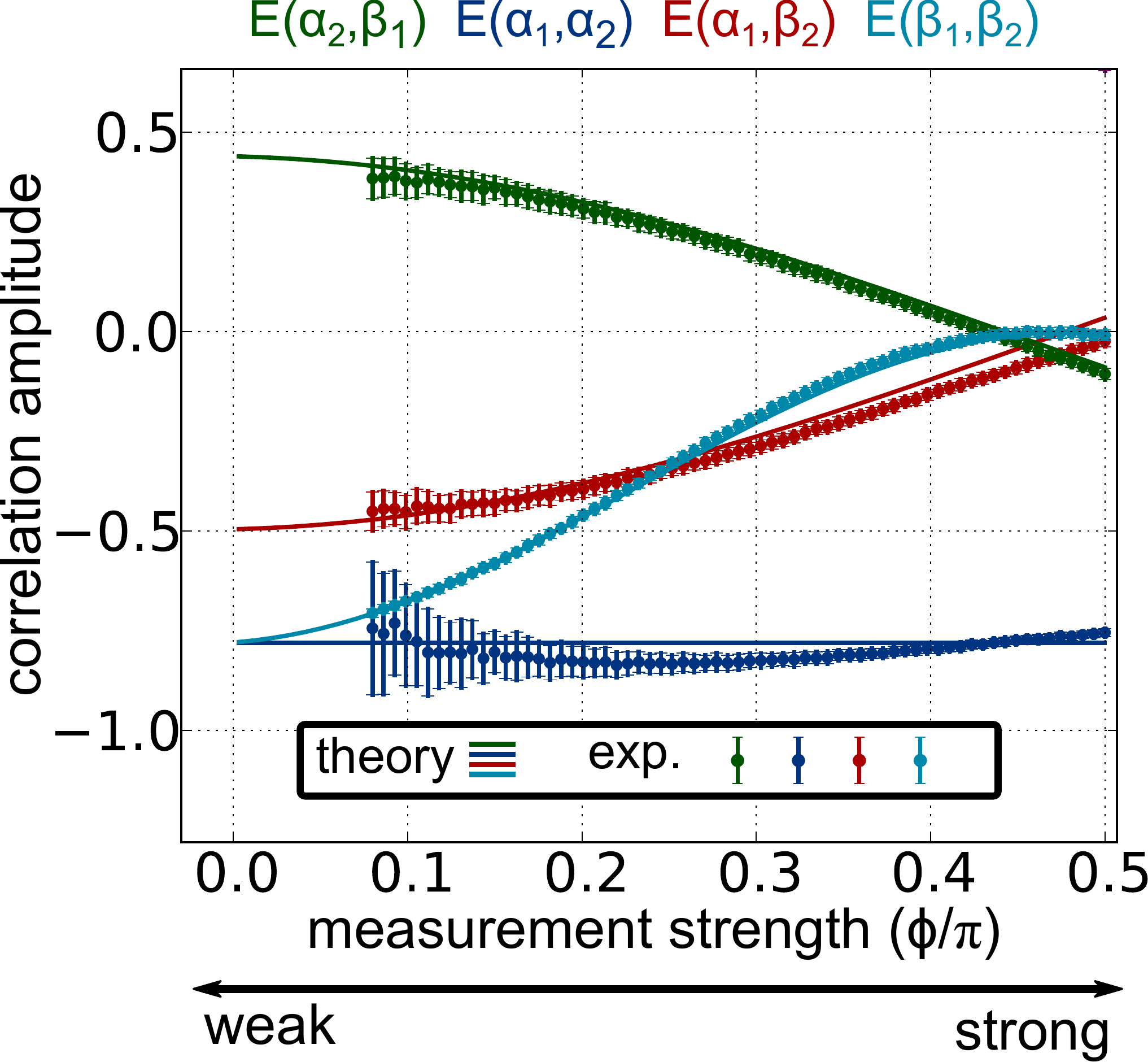}
    \caption{Graph showing both experimental data (points) and theoretical predictions (lines) for the four BLGI terms vs. measurement strength $\phi$.  The data set was taken by averaging together 200 hundred traces in which each point was measured 3000 times for a total of 600,000 iterations per point.  The error bars represent 10 standard deviations of the mean to demonstrate the scaling the ancilla measurement noise vs. measurement strength.  Distinct differences in the behavior arise from experimental imperfections and their depedence on different qubit correlations.  In particular the difference between $E(\beta_1,\beta_2)$ and $E(\alpha_1,\beta_2)$ or $E(\beta_1,\alpha_2)$ highlights the dependence of entanglement collapse on the measurement strength of both ancilla qubits.}
    \label{fig:blgiTerms}
\end{figure}

During the numeric optimization of the pulse sequence, single qubit phases can be adjusted slightly to increase correlations, leading to a larger violation.  This is equivalent to changing the final rotation angle of the detectors slightly ($\sim 3$ degrees). The nature of the BLGI makes it immune to such rotations as loss of correlations in one correlator is naturally made up for in another.  additionally, the initial detector rotation $b$ was chosen based on the maximum of the original CHSH measurements.  Due to differences in qubit coherence this does not necessarily occur at $\pi/4$, but at a slightly smaller angle.  The individual BLGI correlator terms measured in this experiment along with theory curves accounting for these realistic rotations are plotted in Fig.\,\ref{fig:blgiTerms}.  The behavior of each individual term depends on the type of qubits being correlated.  The term $E(\alpha_1,\alpha_2)$ holds roughly constant, close to the expected value of $-1/\sqrt{2}$.  $E(\alpha_1,\beta_2)$ and $E(\beta_1,\alpha_2)$ start close to zero, and converge to around $\pm0.5$.  The behavior of $E(\beta_1,\beta_2)$ best matches expectations. It begins at 0 for strong ancilla measurement and converges near $1/\sqrt{2}$ at perfectly weak measurement, following a slightly s-shaped curve.

\subsection{Sample fabrication and characterization}

Devices are fabricated identically to Ref.~\cite{barends:gates}, and extensive calibration was documented in Ref.~\cite{9xMon}.

\subsection{Device parameters}

The device parameters are listed in table~\ref{tab:params}. Note
that the coupling rate $g$ is defined such that strength of the level
splitting on resonance (swap rate) is $2g$ (Ref.~\cite{swapnote}).

\begin{table*}[t!]
\caption{Parameters for the device when running the BLGI algorithm. $f$ are frequencies. $\eta$ is qubit nonlinearity. $g$ is coupling strength. $\kappa$ is resonator leakage rate. }
\begin{tabular}{l c c c c }
  \hline
  \hline
   & $Q_0$ & $Q_1$ & $Q_2$ &$ Q_3$ \\
   \hline
   \multicolumn{5}{c}{\textbf{Qubit frequencies and coupling strengths}}\\
   $f_{10}^{max}$ (GHz) & 5.30 & 5.93 & 5.39 & 5.90 \\
   $\eta/2\pi$ (GHz) & -0.230 & -0.216 & -0.229 & -0.214 \\
   $f_{10}^{idle}$ (GHz)&  4.53 & 5.42 & 4.67 & 5.55 \\
   $f_\mathrm{res}$ (GHz) & 6.748 & 6.626 & 6.778 & 6.658 \\
   $g_\mathrm{res}/2\pi$ (GHz) & 0.110 & 0.128 & 0.111 & 0.109 \\
   $g_\mathrm{qubit}/2\pi$ (MHz) & \multicolumn{2}{c}{13.8} & \multicolumn{2}{c}{14.1} \\
   $g_\mathrm{qubit}/2\pi$ (MHz) & & \multicolumn{2}{c}{14.5}\\
   $1/\kappa_\mathrm{res}$ (ns) & 675 & 69 & 555 & 30 \\
   \hline
   \multicolumn{5}{c}{\textbf{Readout (RO) parameters}}\\
   RO error & 0.015 & 0.004 & 0.067 & 0.007 \\
   Thermal $\ket{1}$ pop. & 0.013 & 0.007 & 0.028 & 0.01 \\
   RO pulse length (ns) & 1000 & 300 & 1000 & 300 \\
   RO demodulation length (ns) & 1000 & 300 & 1000 & 300 \\
   \hline
   \multicolumn{5}{c}{\textbf{Qubit lifetime at idling point}}\\
   $T_1$ ($\mu$s) & 26.3 & 24.7 & 39.2 & 21.3 \\
  \hline
  \hline
\label{tab:params}
\end{tabular}
\end{table*}

\newpage

\addcontentsline{toc}{chapter}{Bibliography}
\bibliographystyle{naturemag}
\bibliography{references}